\documentclass[12pt,aps,prd,superscriptaddress,floatfix,nofootinbib,longbibliography]{revtex4-1}

\usepackage[utf8]{inputenc}
\usepackage[T1]{fontenc}
\usepackage{graphicx}
\usepackage{color}
\usepackage{amsmath}
\usepackage{amssymb}
\usepackage{hyperref}
\usepackage{wrapfig}
\newcommand{\pT} {p_{\mathrm{T}}}
\newcommand{\nch} {N_{\mathrm{ch}}}
\newcommand{\npart} {N_{\mathrm{part}}}
\newcommand{\sqrtnn} {\sqrt{s_{\mathrm{NN}}}}
\newcommand{\lr}[1]{\left\langle #1\right\rangle}

\newcommand{\trento}{T$\mathrel{\protect\raisebox{-2.1pt}{R}}$ENTo}

\usepackage{tikz}
\usetikzlibrary{tikzmark}

\begin{document}

\title{Imaging the initial condition of heavy-ion collisions\\ and nuclear structure across the nuclide chart}
\newcommand{\LBNL}{Lawrence Berkeley National Laboratory, Berkeley, CA 94720, USA}
\newcommand{\MSU}{Michigan State University, East Lansing, MI 48824, USA}
\newcommand{\UIUC}{University of Illinois at Urbana-Champaign, IL 61801, USA}
\newcommand{\BNL}{Brookhaven National Laboratory, Upton, NY 11973, USA}
\newcommand{\RBRC}{RIKEN BNL Research Center, Brookhaven National Laboratory, Upton, NY 11973, USA}
\newcommand{\SBU}{Stony Brook University, Stony Brook, NY 11794, USA}
\newcommand{\RICE}{Rice University, Houston, TX 77005, USA}
\newcommand{\MIT}{Massachusetts Institute of Technology, Cambridge, MA 02139, USA}
\newcommand{\UH}{University of Houston, Houston, TX 77204, USA}
\newcommand{\OSU}{The Ohio State University, Columbus, OH 43210-1117, USA}
\newcommand{\WSU}{Wayne State University, Detroit, MI 48201, USA}
\newcommand{\ORNL}{Oak Ridge National Laboratory, Oak Ridge, TN, 37830, USA}
\newcommand{\HeidelbergU}{Institut f\"{u}r Theoretische Physik, Universit\"{a}t Heidelberg, Philosophenweg 16, 69120 Heidelberg, Germany}
\newcommand{\USPaulo}{Instituto de F\'isica, Universidade de S\~ao Paulo, R. do Mat\~ao 1371, 05508-090 S\~ao Paulo, Brazil}
\newcommand{\CERN}{CERN, CH-1211 Genève 23, Switzerland}
\newcommand{\NBI}{Niels Bohr Institute, University of Copenhagen, Blegdamsvej 17, 2100 Copenhagen, Denmark}
\newcommand{\SA}{ IRFU, CEA, Université Paris-Saclay, 91191 Gif-sur-Yvette, France}
\newcommand{\KENT}{Physics Department, Kent State University, Kent, OH 44242, USA}
\newcommand{\FUDAN}{Key Laboratory of Nuclear Physics and Ion-beam Application (MOE), Institute of Modern Physics, Fudan University, Shanghai 200433, China}

\author{Jiangyong Jia}\email[Correspond to\ ]{jiangyong.jia@stonybrook.edu}\affiliation{\SBU}\affiliation{\BNL}
\author{Giuliano Giacalone}\email[Correspond to\ ]{giulianogiacalone@gmail.com}\affiliation{\CERN}
\author{Benjamin Bally}\affiliation{\SA}
\author{James Daniel Brandenburg}\affiliation{\OSU}
\author{Ulrich Heinz}\affiliation{\OSU}
\author{Shengli Huang}\affiliation{\SBU}
\author{Dean Lee}\affiliation{\MSU}
\author{Yen-Jie Lee}\affiliation{\MIT}
\author{Constantin Loizides}\affiliation{\RICE}
\author{Wei Li}\affiliation{\RICE}
\author{Matthew Luzum}\affiliation{\USPaulo}
\author{Govert Nijs}\affiliation{\CERN}
\author{Jacquelyn Noronha-Hostler}\affiliation{\UIUC}

\author{Mateusz Ploskon}\affiliation{\LBNL}
\author{Wilke van der Schee}\affiliation{\CERN}
\author{Bjoern Schenke}\affiliation{\BNL}
\author{Chun Shen }\affiliation{\WSU}\affiliation{\RBRC}
\author{Vittorio Som\`a}\affiliation{\SA}
\author{Anthony Timmins}\affiliation{\UH}
\author{Zhangbu Xu}\affiliation{\KENT}\affiliation{\BNL}
\author{You Zhou}\affiliation{\NBI}

\maketitle
\section*{Abstract}
High-energy nuclear collisions encompass three key stages: the structure of the colliding nuclei, informed by low-energy nuclear physics, the \textit{initial condition}, leading to the formation of quark-gluon plasma (QGP), and the hydrodynamic expansion and hadronization of the QGP, leading to final-state hadron distributions that are observed experimentally. Recent advances in both experimental and theoretical methods have ushered in a precision era of heavy-ion collisions, enabling an increasingly accurate understanding of these stages. However, most approaches involve simultaneously determining both QGP properties and initial conditions from a single collision system, creating complexity due to the coupled contributions of these stages to the final-state observables.

To avoid this, we propose leveraging established knowledge of low-energy nuclear structures and hydrodynamic observables to independently constrain the QGP's initial condition. By conducting comparative studies of collisions involving isobar-like nuclei -- species with similar mass numbers but different ground-state geometries -- we can disentangle the initial condition's impacts from the QGP properties. This approach not only refines our understanding of the initial stages of the collisions but also turns high-energy nuclear experiments into a precision tool for imaging nuclear structures, offering insights that complement traditional low-energy approaches.

Opportunities for carrying out such comparative experiments at the Large Hadron Collider (LHC) and other facilities could significantly advance both high-energy and low-energy nuclear physics. Additionally, this approach has implications for the future Electron-Ion Collider (EIC).  While the possibilities are extensive, we focus on selected proposals that could benefit both the high-energy and low-energy nuclear physics communities. Originally prepared as input for the long-range plan of U.S. nuclear physics, this white paper reflects the status as of September 2022, with a brief update on developments since then.

\newpage
\section{Introduction: from nuclear structure to heavy-ion collisions}
Collective behavior in many-body systems governed by the strong nuclear force emerges ubiquitously across energy scales, and plays an instrumental role in our understanding of the phenomenology of such complex systems. In the zero temperature realm of atomic nuclei, strong collective correlations of nucleons lead to a range of fascinating structure properties, such as the emergence of rotational bands, which are naturally explained via notions of nuclear deformations and fluctuating intrinsic nuclear shapes \cite{bohr}. At high temperatures, nucleons melt into fundamental constituents, quarks and gluons, to form the so-called quark-gluon plasma (QGP), whose collective description in terms of fluid dynamics has enabled us to explain a wealth of experimental data from high-energy nuclear collisions \cite{Teaney:2009qa,Busza:2018rrf}. 

Recently, collisions of ions of similar mass at high energy, such as in the BNL RHIC isobar run of $^{96}$Ru+$^{96}$Ru and $^{96}$Zr+$^{96}$Zr collisions,  have led to the experimental demonstration of the direct impact of structural properties of nuclei on the collective flow of the produced QGP \cite{STAR:2021mii}. Enabling such a connection is the fact that high-energy collisions probe, on an event-by-event basis, nucleon configurations from collapsed nuclear wave functions in the overlap region \cite{Miller:2007ri,Shou:2014eya,Giacalone:2020ymy}. This is made possible by the ultra-short time duration for the interaction between the two ions at high energy. Sensitivity to individual realizations of nucleon configurations, combined with the large number of particles produced in each high-energy collision (up to 30,000 particles in a Pb+Pb collision at CERN LHC energy~\cite{ALICE:2016fbt}) enables a direct link between multi-particle correlations in the final state of the collisions and multi-nucleon correlations in the colliding nuclear states.
The way high-energy collisions of nuclei access the nuclear structure is, therefore, akin to the techniques employed in the study of many-body correlations in highly-controllable quantum systems, such as cold atom gases \cite{jochim1,jochim2}, where the coordinates of individual constituents are measured via imaging techniques. High-energy collisions are the ideal tool for imaging the collective structure of atomic nuclei, as opposed to electron-nucleus scattering, where more local information about parton structure or short-range nucleon correlations is accessible. 

A major research goal in high-energy nuclear physics is the characterization of the QGP in terms of medium properties, such as specific shear and bulk viscosities, $\eta/s$ and $\zeta/s$, or the jet quenching transport parameter, $\hat q$~\cite{Busza:2018rrf}. The precision achievable in this characterization, e.g., in state-of-the-art Bayesian analyses~\cite{Bernhard:2016tnd,JETSCAPE:2020xug,Nijs:2020ors,Xie:2022ght,JETSCAPE:2022ixz}, is impacted by our uncertain knowledge of the mechanism of energy deposition in the interaction of two nuclei. Assessing the role of the nuclear structure input will, therefore, reduce this uncertainty, permitting global analyses of data to infer cleaner information about the collision dynamics, and in turn the knowledge of the QGP initial condition~\cite{Giacalone:2021uhj}. Conversely, a major direction of research in nuclear structure theory focuses on the emergence of nuclear properties from fundamental theory \cite{Hergert:2020bxy}. Such \textit{ab initio} approaches aim at describing strongly-correlated nuclear systems from approximate (yet systematically improvable) solutions of the Schrödinger equation with nucleon-nucleon and three-nucleon interactions constructed in an effective theory of low-energy QCD. These efforts find a natural application in the phenomenology of multi-particle correlations in high-energy nuclear collisions. Once the response of the QGP initial condition to nuclear structure is established, one could use measurements in heavy-ion collisions to test the results of \textit{ab initio} approaches in a way that is complementary to low-energy experiments. The systematic use of \textit{ab initio} results as an input for the model building of nuclear collisions will then permit us to assess, in particular, the consistency of nuclear phenomena across energy scales.

Given the rapid progress in the development of \textit{ab initio} theories of nuclear structure, and considering that the nuclear program at the CERN LHC in the next decade is largely to be defined, it is timely to identify the physics opportunities based on the synergy of these two areas from which the nuclear community as a whole could benefit.

\section{Manifestation of nuclear structure in high-energy nuclear collisions}

\subsection{Methodology}\label{sec:2.1}

\begin{figure}[t]
\includegraphics[width=0.85\linewidth]{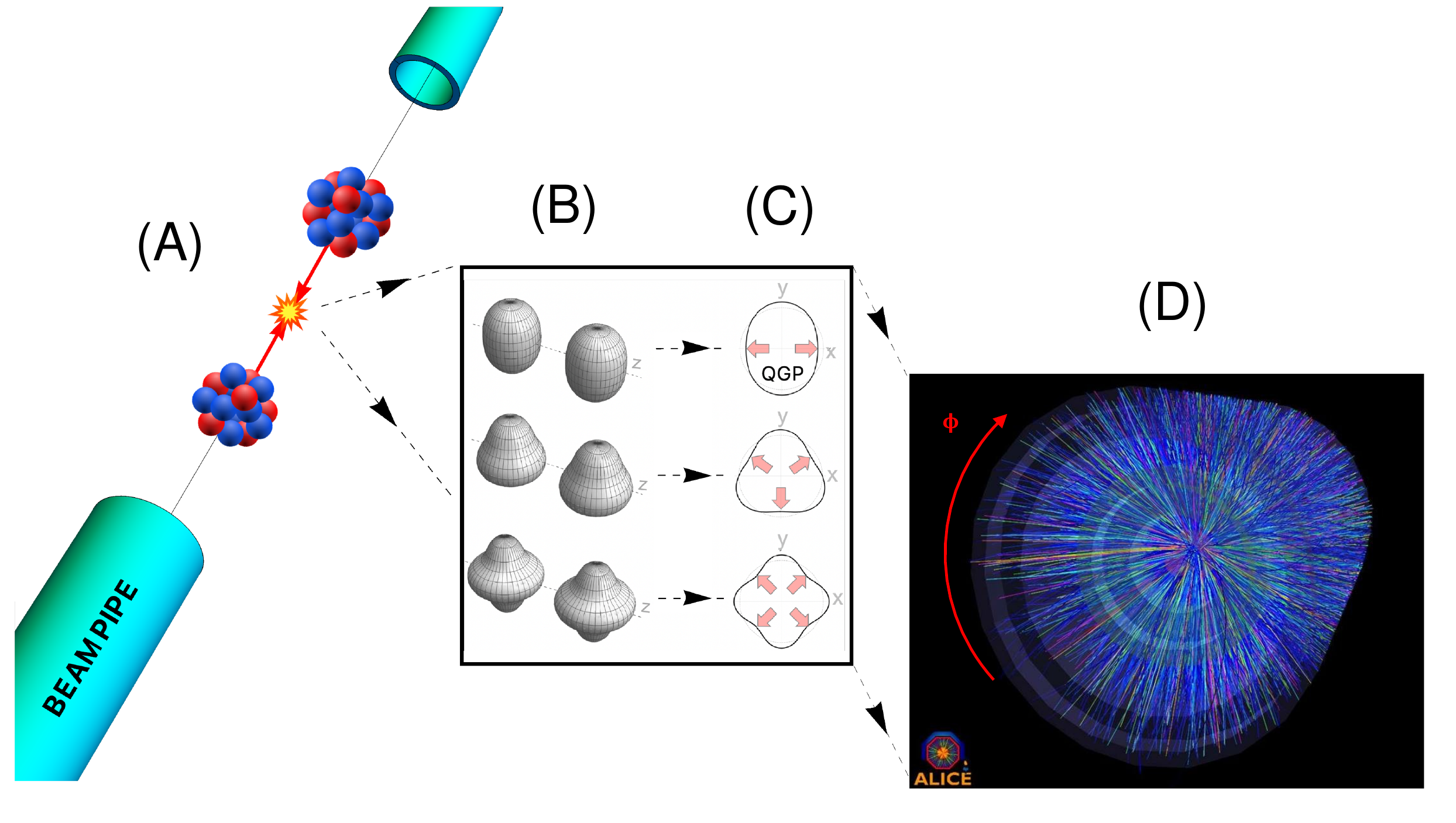}
\caption{\label{fig:1} Schematic view of a relativistic heavy-ion collision, highlighting the role played by the collective properties of the colliding ions in shaping the geometry of the produced quark-gluon plasma (QGP). The Lorentz contraction of the two nuclei in the $z$-direction, by factor $\gamma\sim100$ at the BNL RHIC or over 1000 at the CERN LHC, is not shown. See text for a detailed description.}
\end{figure}

Figure~\ref{fig:1} illustrates the method for accessing the structure of ions colliding at relativistic energies. (A) Two nuclei are smashed in a high-energy collider (the large Lorentz contraction in the beam direction is not shown). (B) At the time of interaction, the nuclei are characterized by nontrivial geometries of nucleon configurations, including deformations and radial profiles. (C) The geometry of such configurations is reflected in the initial condition of the created QGP. The subsequent hydrodynamic expansion of this system, driven by pressure-gradient forces, converts the spatial asymmetries in the initial shape into the momentum asymmetries of emitted particles in the transverse plane.  (D) Experimentally, transverse momentum ($\pT$) asymmetries can be revealed via a Fourier expansion of the particle distributions in azimuthal angle:
\vspace*{-0.2cm}\begin{equation}\label{eq:1}
\frac{d^2N}{d\pT d\phi} = \frac{dN}{2\pi d\pT} \left (1 + 2 \sum_{n=1}^{\infty} v_n \cos n(\phi-\Phi_n) \right ),
\end{equation}
where the Fourier harmonics $V_n = v_n e^{in\Phi_n}$ are coefficients of anisotropic flow. The most significant harmonics are $V_2$, elliptic flow, reflecting the elliptical asymmetry of the geometry of the QGP, and $V_3$, reflecting the triangular asymmetry~\cite{Ollitrault:1992bk,Alver:2010gr,Teaney:2010vd}. We note that the total particle multiplicity, $\nch = \int d\pT dN/d\pT$, is proportional to the amount of energy deposited in a collision, which in turn is determined by the number of nucleons, $\npart$, participating in the interaction. The slope of the $\pT$ spectra reflects the strength of the radial expansion, characterized by $\lr{\pT} = \frac{1}{\nch} \int dp_{\rm T} ~p_{\rm T} dN/dp_{\rm T}$, which is inversely related to the transverse size of the overlap region~\cite{Bozek:2012fw}. Due to these relations, inherent to the hydrodynamic description, information about the structure of the colliding ions can be inferred from the detected final-state particles.

\begin{figure}[t]
\includegraphics[width=0.95\linewidth]{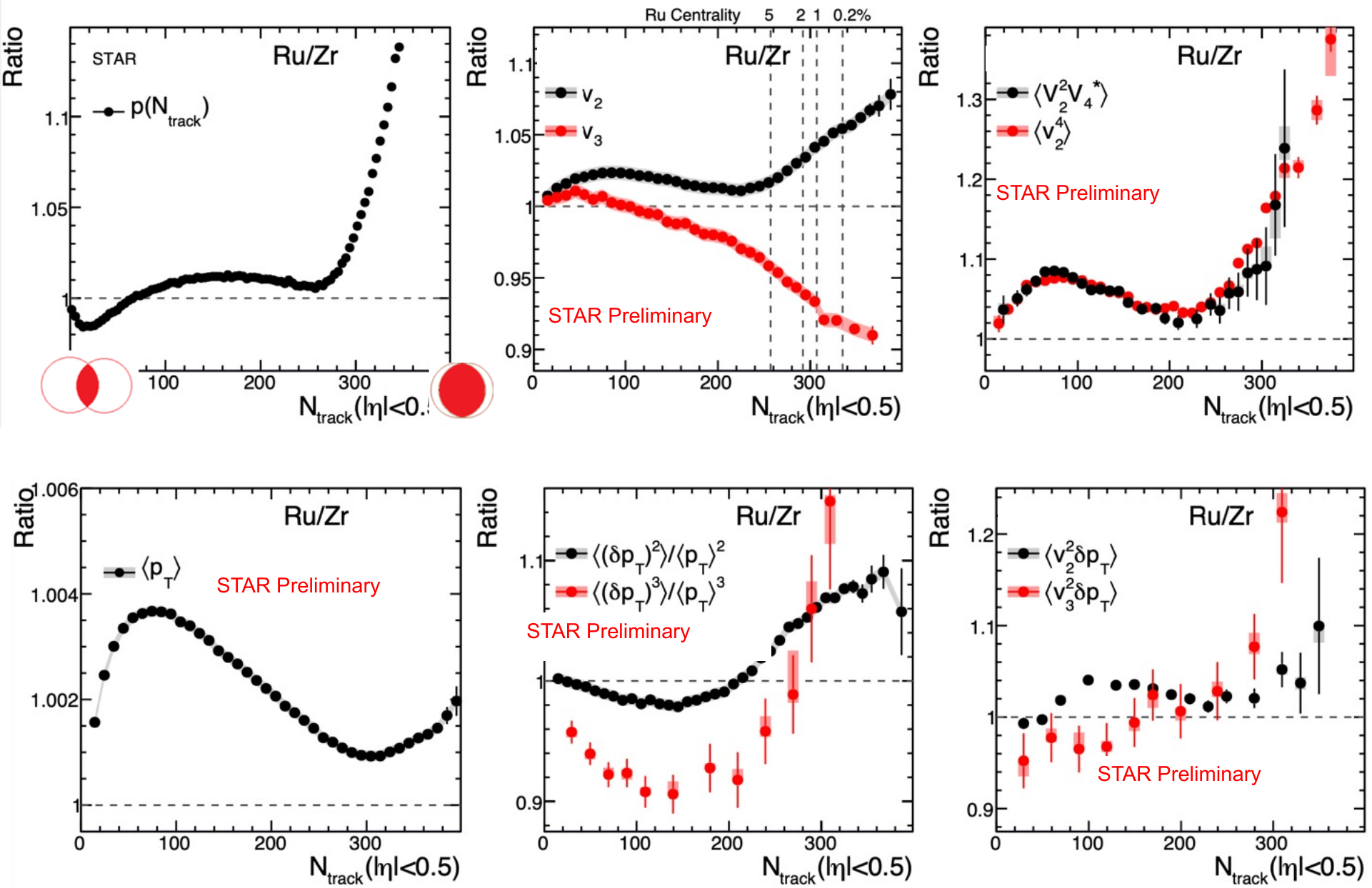}
\caption{\label{fig:2} Ratios of observables taken between $^{96}$Ru+$^{96}$Ru and $^{96}$Zr+$^{96}$Zr collisions as a function of $\nch$, as measured by the STAR Collaboration (Preliminary results). A total of ten ratios are shown.}
\end{figure}

The most direct way of observing the impact of the nuclear structure via this method is through comparing observables measured in collisions of species that are close in mass. Isobars, i.e., nuclides having the same mass number, are ideal candidates for such studies \cite{Giacalone:2021uhj}, as explicitly demonstrated by experimental data from $^{96}$Zr+$^{96}$Zr and $^{96}$Ru+$^{96}$Ru collisions, collected in 2018 at the BNL RHIC and released three years later \cite{STAR:2021mii}. Given two isobars, $X$ and $Y$, and a given observable, $\mathcal{O}$, we ask the following question:
\begin{equation}
\boxed{
\label{eq:2}
\frac{\mathcal{O}_{X+X} }{\mathcal{O}_{ Y+Y } } \stackrel{?}{=} 1
}
\end{equation}
Model studies have established that any visible departure from unity in the ratio must originate from differences in the structure of the isobars. In the measurements of the STAR collaboration, structure influences are ubiquitously found. Ratios of more than ten observables taken between $^{96}$Zr+$^{96}$Zr and $^{96}$Ru+$^{96}$Ru have been measured, all displaying distinct and centrality-dependent deviations from unity, as reported in Fig.~\ref{fig:2}~\cite{chunjianhaojie}.  Such rich and versatile information can provide a new type of constraint on the structure of these isobars, as also predicted by early model investigations, which we discuss below.  

\subsection{Nuclear structure input}\label{sec:2.2}
The hydrodynamic model of heavy-ion collisions successfully reproduces a vast set of experimental measurements at the BNL RHIC and the CERN LHC \cite{Nijs:2022rme}. The input to hydrodynamic simulations is the event-by-event distribution of nucleons in the colliding ions. Motivated by low-energy nuclear physics, a Woods-Saxon profile with a nuclear surface expanded in spherical harmonics is routinely employed,
\begin{equation}\label{eq:3}
\rho(r,\theta,\phi) \propto \frac{1}{1+e^{[r-R_0 ( 1 +\beta_2 ( \cos \gamma Y_2^0(\theta,\phi) + \sin \gamma Y_2^2(\theta,\phi)) + \beta_3 Y_3^0(\theta,\phi))]/a_0}},
\end{equation}
where $R_0$ is the half-density radius, $a_0$ is the surface diffuseness, $\beta_2$ is the magnitude of the quadrupole deformation, $\gamma$ determines the relative length of the three axes of the ellipsoid, and $\beta_3$ is the (axial) octupole deformation parameter, where $\beta_3\neq0$ implies a breaking of parity symmetry in the intrinsic nuclear shape.  Hydrodynamic simulations show that any changes in these parameters leave characteristic and detectable impacts on experimental observables such as those shown in Fig.~\ref{fig:2} \cite{Jia:2021oyt,Nijs:2021kvn}. 

Alternatively, hydrodynamic simulations can take pre-sampled nucleon configurations from \textit{ab initio} calculations as input (See, e.g., Refs.~\cite{Lim:2018huo,Rybczynski:2019adt,Summerfield:2021oex,Nijs:2021clz} for such applications in $^{16}$O collisions). Here, diffuseness and deformations emerge directly from many-nucleon correlations in the sampled wave functions. Given the expected rapid progress in the reach and quality of \textit{ab initio} calculations over the next few years~\cite{Hergert:2020bxy,Furnstahl:2021rfk}, this alternative approach should become broadly adopted in the modeling of heavy-ion collisions in future. Full exploitation of such predctions of state-of-the-art nuclear theory  will demonstrate further the scientific relevance of the connection between high-energy observations and low-energy theories.

\subsection{Signatures of intrinsic nuclear shapes}\label{sec:2.3}
\begin{figure}[t]
\includegraphics[width=\linewidth]{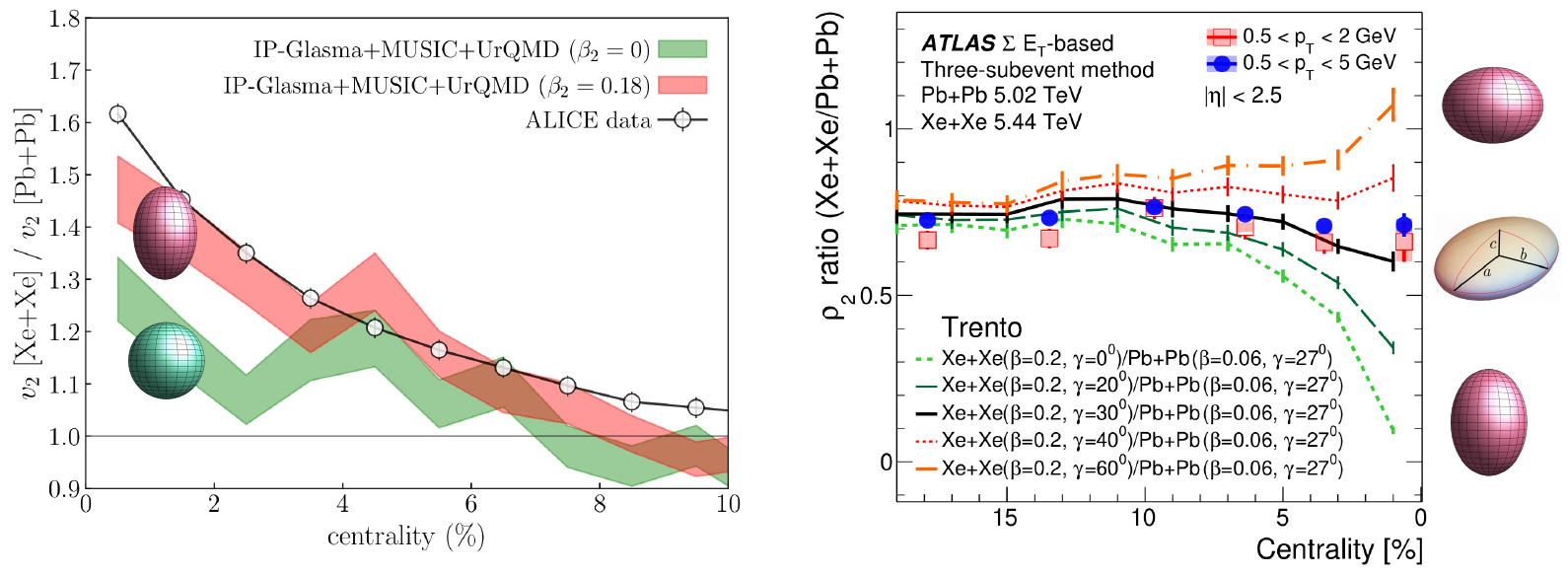}
\caption{\label{fig:3} Modification of multi-particle correlation observables in Xe+Xe collisions compared to the baseline with spherical nuclei provided by Pb+Pb collisions. Left: elliptic flow, $v_2$~\cite{ALICE:2018lao}. Right: correlation between elliptic flow and the average transverse momentum, $\rho_2$~\cite{ATLAS:2022dov}.}
\end{figure}

A crucial observable in high-energy heavy-ion collisions is the rms flow coefficient, $v_n = \sqrt{\langle |V_n|^2 \rangle}$. Numerical and semi-analytical studies show that, for collisions at a given multiplicity (or centrality), $v_n$ is enhanced by the presence of nuclear deformations in the colliding ions, following\cite{Giacalone:2018apa,Giacalone:2021uhj,Giacalone:2021udy,Jia:2021tzt},
\vspace*{-0.3cm}\begin{equation}
\label{eq:4}
    v_n^2 \approx b_0 + b_1 \beta_n^2,
\end{equation}
where $b_0$ and $b_1$ are positive coefficients that depend on centrality. The enhancement predicted by Eq.~\eqref{eq:4} would show up, in particular, when comparing collisions of deformed nuclei to collisions of spherical nuclei. A powerful way to do so is to compare deformed ion collisions with collisions of nearly spherical $^{208}$Pb ions. The left panel of Fig.~\ref{fig:3} reveals an enhanced $v_2$ in $^{129}$Xe+$^{129}$Xe collisions compared to $^{208}$Pb+$^{208}$Pb collisions \cite{Giacalone:2017dud}, as observed by the ALICE collaboration \cite{ALICE:2018lao}. A state-of-the-art calculation~\cite{Schenke:2020mbo} confirms the origin of this effect due to the large $\beta_2$ of $^{129}$Xe.

Concerning the triaxiality, $\gamma$ in Eq.~\eqref{eq:1}, revealing its presence requires the use of three-particle correlations. The most sensitive observable is the correlation of the shape of the QGP with its size \cite{Giacalone:2019pca}, measurable experimentally via a correlation between $v_n^2$ and the fluctuation of transverse momentum, $\delta\pT=\pT-\lr{\pT}$, at a given multiplicity. This quantity is conveniently formulated as a Pearson coefficient \cite{Bozek:2016yoj}, $\rho_n = \frac{\lr{v_n^2\delta\pT}}{\sqrt{\left(\lr{v_n^4}-\lr{v_n^2}^2\right)}\sqrt{\lr{\delta \pT\delta \pT}}}$. For quadrupole deformation, theoretical work shows the following leading-order dependence \cite{Jia:2021qyu}
\begin{equation}\label{eq:4b}
\rho_2 \approx b_0' - b_1' \beta_2^3 \cos ( 3 \gamma),
\end{equation}
where $b_0'$ and $b_1'$ are positive coefficients.  In the presence of large $\beta_2$, moving from oblate ($\gamma=60^\circ$) to prolate ($\gamma=0$) shapes decreases $\rho_2$ in a substantial way. A recent measurement at RHIC shows precisely $\rho_2<0$ in central U+U collisions \cite{jia}, which is explained naturally by the large prolate deformation of $^{238}$U \cite{Giacalone:2020awm}, $\beta_2\sim0.28$, $\gamma=0$. The nucleus $^{129}$Xe is particularly interesting for such a study, as its shape is considerably deformed and also triaxial, $\beta_2=0.2$ and $\gamma\approx30^\circ$ \cite{Vietze:2014vsa,Bally:2021qys,Bally:2022rhf}. In the right panel of Fig.~\ref{fig:3}, model calculations assuming oblate, triaxial, and prolate $^{129}$Xe shape show a strong modification of $\rho_2$ in $^{129}$Xe+$^{129}$Xe collisions with respect to the $^{208}$Pb+$^{208}$Pb collisions \cite{Bally:2021qys}. Measurements from the ATLAS collaboration indeed confirm the triaxial scenario \cite{ATLAS:2022dov}. One important point is that the combined use of $v_2^2$ and $\rho_2$ can simultaneously constrain $\beta_2$ and $\gamma$.

In the octupole sector, much less is known from low-energy physics \cite{Butler:1996zz}. Direct evidence of octupole deformation in excitation bands of atomic nuclei is scarce, because octupole deformation rarely manifests as a mean-field effect (\textit{static} deformations) \cite{Robledo:2011nf,Cao:2020rgr}, as in a simple rotor model. However, \textit{dynamical} octupole correlations at the beyond-mean-field level are present in essentially all nuclei \cite{Robledo:2014nia}, and should leave their fingerprint in the nucleon configurations from \textit{ab initio} calculations. 
High-energy nuclear collisions, probing configurations of nucleons on an event-by-event basis, give access to all such non-static deformations in the ground states in the same way as the static ones.

\begin{figure}[t]
\includegraphics[width=0.5\linewidth]{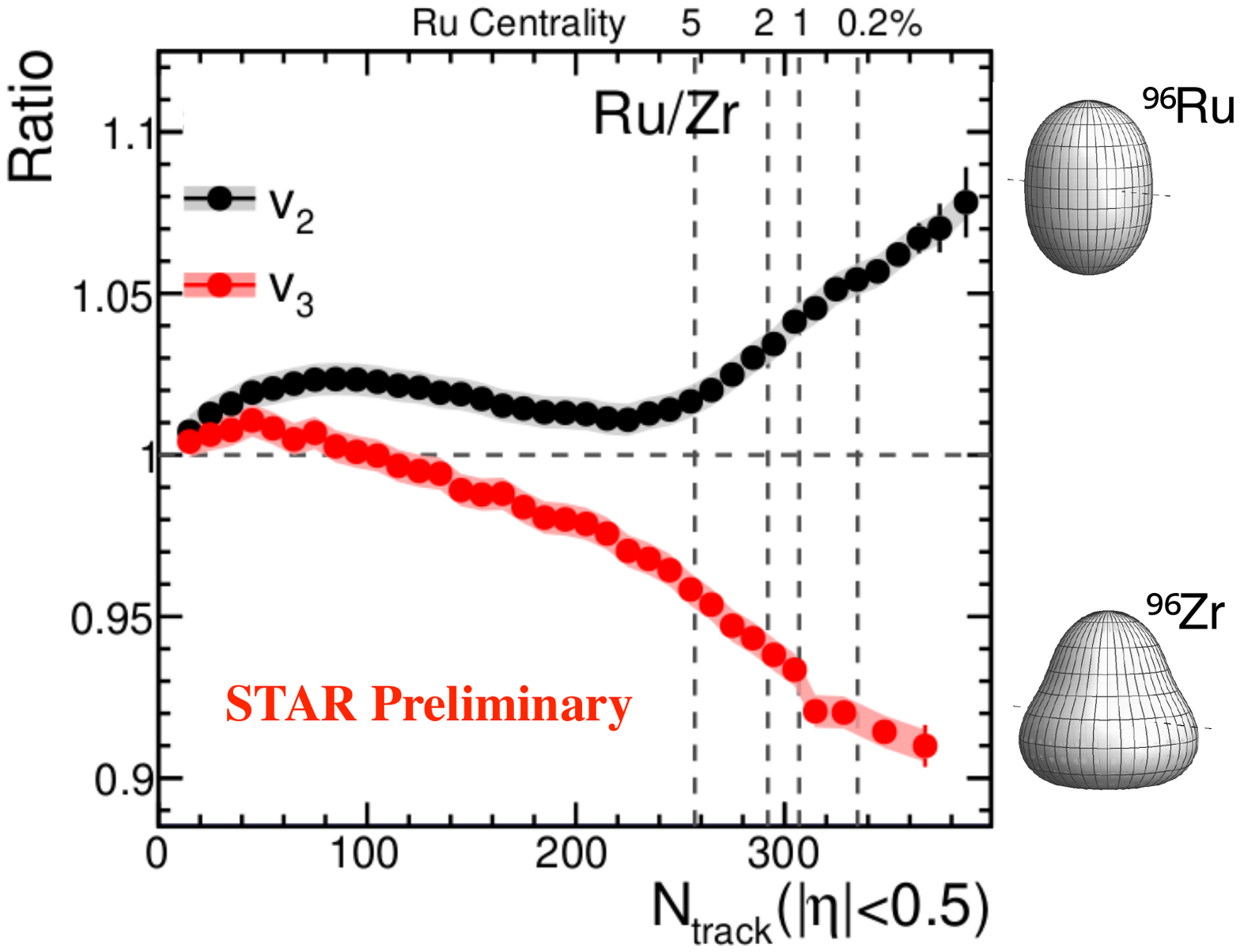}
\caption{\label{fig:4} Preliminary ratios of flow coefficients, $v_n$, taken between Ru+Ru and Zr+Zr collisions. The suppression of the $v_3$ ratio at large multiplicity is due to an enhancement of $v_3$ in Zr+Zr collisions.}
\end{figure}

One of the breakthrough outcomes of the isobar collision campaign at RHIC is reported in Fig.~\ref{fig:4}, also shown in the mid-bottom panel of Fig.~\ref{fig:2}. The ratio of $v_n$ taken between Ru+Ru and Zr+Zr collisions shows significant departures from unity. The data implies that $^{96}$Ru has a larger $\beta_2$ than $^{96}$Zr, as expected from low-energy experiments. A similar departure for $n=3$, showing an enhanced $v_3$ in Zr+Zr collisions, can only be ascribed to $^{96}$Zr having a sizable $\beta_3$ \cite{Zhang:2021kxj}, which is not predicted by mean-field energy density functional calculations \cite{Cao:2020rgr,Rong:2022qmu}. The results of the STAR collaboration demonstrate that heavy-ion collisions offer a clean access route to multi-nucleon correlations that are both difficult to quantify from traditional low-energy experiments and hard to predict from phenomenological models.

\subsection{Radial profiles and relation to neutron distributions}\label{sec:2.4}
The nuclear radial profile, determined by the $R_0$ and $a_0$ parameters in Eq.~\eqref{eq:2}, influences the area and the density of the overlap region. In general, a smaller $a_0$ or $R_0$ for a fixed mass number leads to a sharper edge in the overlap geometry, leading to a more compact QGP, larger pressure gradients, and hence larger $\lr{\pT}$ and $v_n$. The impact is more significant in off-central collisions where the overlap region is smaller, and sensitivity to a variation in $R_0$ and $a_0$ is larger. Indeed, model studies show that the probability distributions of $\npart$, and hence the distribution of $\nch$, $p(\nch)$, as well as $\lr{\pT}$ and $v_2$, are largely impacted by variations in $a_0$ and $R_0$~\cite{Li:2019kkh,Xu:2021uar,Jia:2021oyt}.

Due to model-dependent systematics, constraining the radial nuclear profile in a single collision system is difficult. Such limitation is largely overcome by comparing experimental observables between systems close in size, such as isobars. Assuming the differences of radial parameters are small, deviation of isobar ratios from unity can be approximated by (taking $^{96}$Ru and $^{96}$Zr as an example)
\begin{equation}
\label{eq:5}
\frac{\mathcal{O}_{{\mathrm Ru}} }{\mathcal{O}_{\mathrm Zr} }\approx 1+ c_0 (R_{0,\rm Ru}-R_{0,\rm Zr})+ c_1 (a_{0,\rm Ru}-a_{0,\rm Zr})\;,\; \mathcal{O}\equiv p(\nch), v_2, \mbox{or}\; \lr{\pT}, 
\end{equation}
where the coefficients $c_0$ and $c_1$ depend only on the mass number at a given centrality or multiplicity and are insensitive to the final state effects~\cite{Zhang:2022fou}. These simple equations describe well the isobar ratios, as verified in recent transport model simulations \cite{Jia:2021oyt}. 

\begin{figure}[t]
\includegraphics[width=1\linewidth]{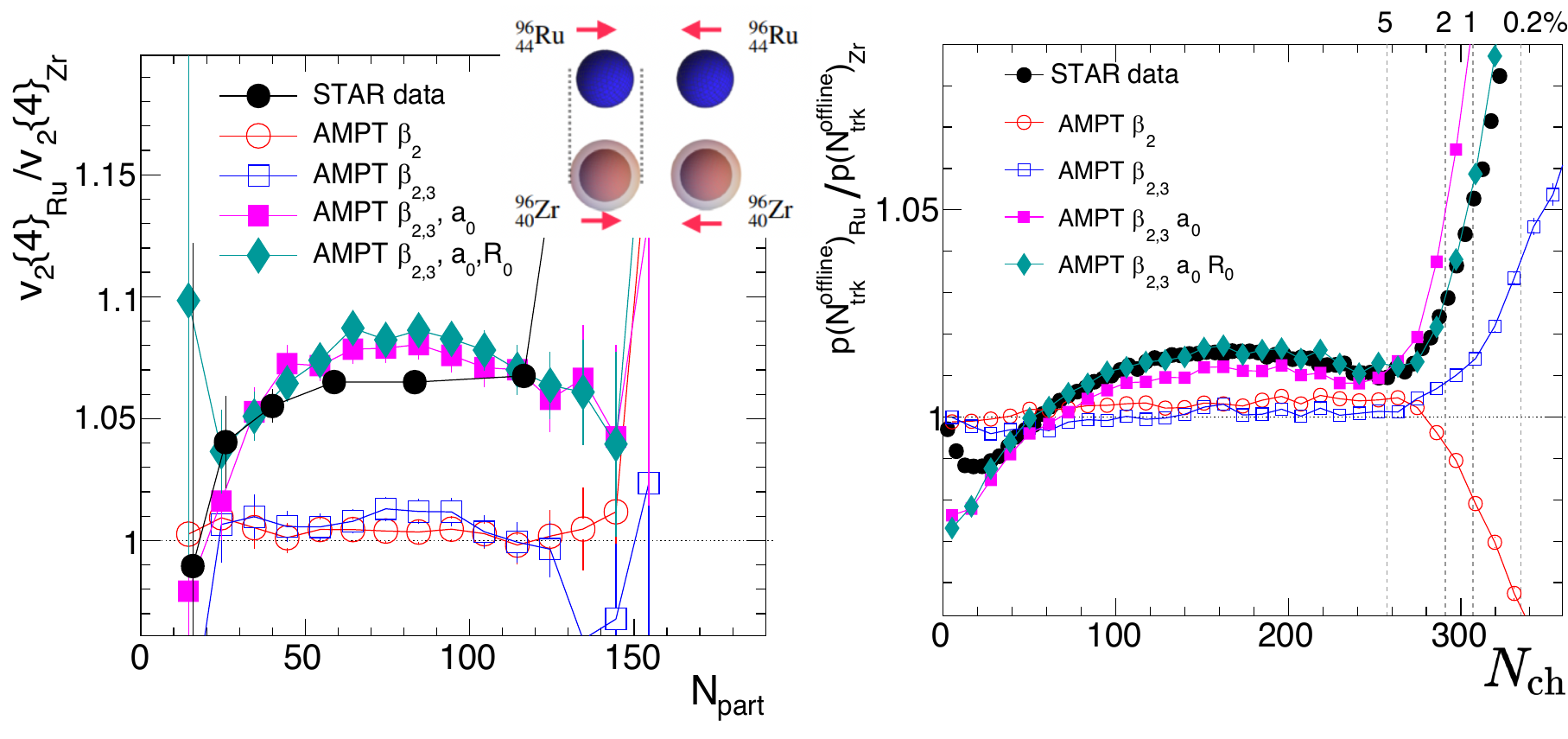}
\caption{\label{fig:5} 
Ratios of observables taken between Ru+Ru and Zr+Zr collisions. The inset in the left panel shows how the neutron excess of $^{96}$Zr compared to $^{96}$Ru yields a more diffuse nuclear surface, i.e., a larger $a_0$ and a slightly smaller $R_0$ in Eq.~\eqref{eq:1}. Left: the impact of the larger $a_0$ of $^{96}$Zr manifests predominantly in the fourth-order cumulant of $v_2$, $v_2\{4\}$, which originates from the intrinsic ellipticity of the QGP due to the finite collision impact parameter. Right: both $a_0$ and $R_0$ differences contribute to a broad hump in the ratio of $p(\nch)$. Calculations are obtained within the A Multi Phase Transport (AMPT) model \cite{Lin:2004en}. }
\end{figure}

Energy density functional calculations suggest that $^{96}$Zr has a larger diffuseness but a smaller radius than $^{96}$Ru, i.e. $\Delta a_0 \equiv a_{0,\rm Ru}-a_{0,\rm Zr}<0$ and  $\Delta R_0 \equiv R_{0,\rm Ru}-R_{0,\rm Zr}>0$~\cite{Li:2018oec,Xu:2021vpn}. As shown in Fig.~\ref{fig:5}, a transport calculation implementing such differences can quantitatively describe the measured ratios of $v_2\{4\}$ (the fourth-order cumulant of elliptic flow fluctuations) and of the distribution of charged particle number, $p(\nch)$. The cumulant $v_2\{4\}$ measures the flow originating from the intrinsic ellipticity acquired by the QGP due to the finite impact parameter of the collisions \cite{Voloshin:2007pc}. Figure~\ref{fig:5} shows that such intrinsic ellipticity is impacted only by $\Delta a_0$ \cite{Jia:2022qgl}, but is insensitive to $\Delta R_0$ and nuclear deformations. On the other hand, $p(\nch)$-ratio is sensitive to both $\Delta a_0$ and $\Delta R_0$ \cite{Li:2018oec,Xu:2021vpn}. A dedicated study finds that the ratio of $\pT$ is also sensitive to both $\Delta a_0$ and $\Delta R_0$ \cite{Xu:2021uar}. Therefore, the measurement of isobar ratios provides several independent determinations on the differences $\Delta R_0$ and $\Delta a_0$, which can be confronted against the predictions of low-energy nuclear structure models.

The knowledge of nucleon distribution, in combination with the well-known proton distribution parameters ${a}_p$ and $R_{0p}$ from low-energy experiments, allows one to probe the difference between the rms radius of neutrons and protons in heavy nuclei, $\Delta r_{np}= R_{n}-R_{p}$, known as the neutron skin. The value of $\Delta r_{np}$ is directly related to the slope of the symmetry energy, dubbed $L$, appearing in the equation of state (EOS) of nuclear matter \cite{Vinas:2013hua}. Determinations of $L$ are intensively pursued at low energy because this parameter plays a crucial role in the stability properties of neutron stars~\cite{Lattimer:2006xb,Li:2008gp}. Isobar ratios in high-energy collisions are expected to probe only the difference in the neutron skin, $\Delta (\Delta r_{np}) = \Delta r_{np,\rm Ru}-\Delta r_{np,\rm Zr}$. 
Assuming Woods-Saxon distributions for protons and nucleons, $\Delta r_{np}$ receives a contribution from both half-radius and surface diffuseness \cite{Jia:2021oyt}:
$\Delta (\Delta r_{np}) \propto (R_0\Delta R_0-R_{0p}\Delta R_{0p})+7/3\pi^2(a\Delta a-a_p \Delta a_p)$.  Therefore, collisions of isobars or, in general, of species of similar mass numbers allow one to access detailed information about radial profiles and neutron skins of nuclei systematically.

\section{Science cases at the intersection of nuclear structure and hot QCD}\label{sec:3}

A window to perform collisions with new ions will be opened in the future at the CERN LHC beyond Run3 (2025) and possibly before the shutdown of the BNL RHIC upon successful completion of the sPHENIX program. About 250 stable isotopes in the nuclear chart could be used systematically for such a purpose. About 140 are found in isobar pairs or triplets, as in Tab.~\ref{tab:1}. Our idea is to select nuclear species that would permit us to 1) probe features of high-energy collisions, in particular their initial condition, by exploiting well-known structural properties, 2) extract structure information of the colliding ions that would complement the effort of low-energy experiments, 3) reveal features of colliding ions that are difficult to access in conventional nuclear structure experiments, but have a significant impact on low-energy models. Continued effort is required to identify species that can maximize the scientific impact for both the hot QCD and the nuclear structure communities. For the moment, we have identified four cases that may lead to discoveries via high-energy experiments. They involve nuclides belonging to the mass regions $A\sim20$, $A\sim40$, $A\sim150$, $A\sim200$. More cases are expected to emerge upon further discussion and model studies (see Section~\ref{sec:3.6} for a brief progress update since 2022). 

\begin{table}[t]
\centering
\begin{tabular}{|c|c|c|c|c|c|c|c|c|c|c|c|}\hline
$A$ & isobars & $A$ & isobars & $A$ & isobars & $A$ & isobars & $A$ & isobars& $A$ & isobars \\\hline
36  & Ar, S     &80  & Se, Kr    & 106 & Pd, Cd  &124 & Sn, Te, Xe & \tikzmark{d1} 148& Nd, Sm &\tikzmark{b1} 174 & Yb, Hf    \\ 
40  & Ca, Ar    &84  & Kr, Sr, Mo& 108 & Pd, Cd  &126 & Te, Xe     & \tikzmark{a1} 150& Nd, Sm \tikzmark{d2}  &176 & ~Yb, Lu, Hf~\\ 
46  & Ca, Ti    &86  & Kr, Sr    & 110 & Pd, Cd  &128 & Te, Xe     & 152& Sm, Gd &180 & Hf, W     \\ 
48  & Ca, Ti    &87  & Rb, Sr    & 112 & Cd, Sn  &130 & Te, Xe, Ba & 154& Sm, Gd &184 & W,  Os\tikzmark{b2}\\ 
50  & ~Ti, V, Cr~ &92  & Zr, Nb, Mo& 113 & Cd, In  &132 & Xe, Ba     & 156& Gd,Dy &186 & W,  Os    \\ 
54  & Cr, Fe    &94  & Zr, Mo    & 114 & Cd, Sn  &134 & Xe, Ba     & 158& Gd,Dy &187 & Re, Os    \\ 
64  & Ni, Zn    &96  & Zr, Mo, Ru& 115 & In, Sn  &136 & Xe, Ba, Ce & 160& Gd,Dy &190 & Os, Pt    \\ 
70  & Zn, Ge    &98  & Mo, Ru    & 116 & Cd, Sn  &138 & Ba, La, Ce & 162& Dy,Er &192 & Os, Pt    \\ 
74  & Ge, Se    &100 & Mo, Ru    & 120 & Sn, Te  &142 & Ce, Nd     & 164& Dy,Er &196 & Pt, Hg    \\ 
76  & Ge, Se    &102 & Ru, Pd    & 122 & Sn, Te  &144 & Nd, Sm     & 168& Er,Yb &198 & Pt, Hg    \\ 
78  & Se, Kr    &104 & Ru, Pd    & 123 & Sb, Te  &\tikzmark{c1}  146 & Nd, Sm  \tikzmark{c2}   & 170& Er,Yb \tikzmark{a2}&204 & Hg, Pb\\\hline
\end{tabular}
\begin{tikzpicture}[overlay,remember picture]
  \draw[line width=0.25mm,red] ([shift={(-0.5ex,2.5ex)}]pic cs:a1) rectangle ([shift={(0.4ex,-.7ex)}]pic cs:a2);
\end{tikzpicture}
\begin{tikzpicture}[overlay,remember picture]
  \draw[line width=0.25mm,red] ([shift={(-.4ex,2.5ex)}]pic cs:b1) rectangle ([shift={(2.5ex,-1.1ex)}]pic cs:b2);
\end{tikzpicture}
\begin{tikzpicture}[overlay,remember picture]
  \draw[line width=0.25mm,blue] ([shift={(-0.5ex,2.5ex)}]pic cs:c1) rectangle ([shift={(0.4ex,-.7ex)}]pic cs:c2);
\end{tikzpicture}
\begin{tikzpicture}[overlay,remember picture]
  \draw[line width=0.25mm,blue] ([shift={(-0.5ex,2.5ex)}]pic cs:d1) rectangle ([shift={(0.4ex,-.5ex)}]pic cs:d2);
\end{tikzpicture}
\caption{\label{tab:1} Pairs and triplets of stable isobars (half-life $>10^8~y$). 141 nuclides are listed. The region marked in red contains large strongly-deformed nuclei ($\beta_2>0.2$). The region marked in blue corresponds to nuclides which may present an octupole deformation in their ground state \cite{Cao:2020rgr}.}
\end{table}

\subsection{Stress-testing small system collectivity with $^{20}$Ne}\label{sec:3.1}

The neon-20 nucleus presents the most extreme ground state of all stable nuclides with $A>10$. It is a strongly-deformed object made of five $\alpha$-clusters in a reflection-asymmetric $\alpha$+$^{16}$O molecular configuration~\cite{ne20b,ne20a,Zhou:2015nza} . In terms of the common quadrupole deformation coefficient, the ground state has $\beta_2\approx0.7$, the highest of all stable ground states. The deformation of this nucleus is so large that its impacts can easily survive the large event-by-event fluctuations associated with sampling a small number of nucleons ($\propto 1/\sqrt{A}$). The extreme geometry of $^{20}$Ne enables us to perform nontrivial tests of the initial-state modeling and the hydrodynamic response in small systems. In particular, one can compare $^{20}$Ne+$^{20}$Ne collisions with collisions of nearly-spherical $^{16}$O nuclei (already recorded at RHIC, and are also planned for 2024 at the LHC). The ratios of observables between the two systems will be largely independent of final state transport properties, and hence directly access the variation in the initial condition caused by nuclear structure differences. Having data from $^{20}$Ne+$^{20}$Ne collisions will maximize the scientific output of the $^{16}$O+$^{16}$O run (and vice versa). On the side of nuclear structure theory, \textit{ab initio} approaches have recently been pushed to describe light systems up to $A\sim 40$ \cite{Hergert:2020bxy}, including $^{20}$Ne \cite{Frosini:2021sxj}. Strong deformations in these approaches emerge from genuine $n$-body (up to $A$-body) correlations in the wavefunction generated by inter-nucleon interactions linked to QCD via an effective field theory. Precise measurement of multi-particle correlations in $^{20}$Ne+$^{20}$Ne collisions will provide novel tests of the effectiveness of such \textit{ab initio} calculations in capturing collective effects in strongly-correlated nuclei. 

As a bonus, while collecting $^{20}$Ne+$^{20}$Ne collisions in collider mode at 7 TeV, one can have the same collisions in fixed-target mode at around 0.07 TeV by injecting a $^{20}$Ne gas in the SMOG system of the LHCb experiment \cite{LHCb:2021ysy}. This would enable a study of the $\sqrt{s_{\rm NN}}$ dependence of the initial condition, longitudinal dynamics and geometry of small systems. 
Another, potentially superior way of imaging the structure of light nuclei is to collide them with heavy spherical nuclei, such as in $^{16}$O+$^{208}$Pb or $^{20}$Ne+$^{208}$Pb~\cite{Broniowski:2013dia,Rybczynski:2017nrx}. The shape of overlap region at small impact parameter directly captures the nucleon distribution in light nuclei. Ratios of observables between $^{16}$O+$^{208}$Pb and $^{20}$Ne+$^{208}$Pb collisions will reveal the shape differences between $^{16}$O and $^{20}$Ne. The main advantage over symmetric $^{16}$O+$^{16}$O and $^{20}$Ne+$^{20}$Ne collisions is that asymmetric ``isobar''+Pb collisions produce much more particles and will have a better centrality resolution~\cite{LHCb:2021ysy}. This idea may be extended to even smaller systems such as $^{8}$Be+$^{208}$Pb and $^{12}$C+$^{208}$Pb~\cite{Broniowski:2013dia,Rybczynski:2017nrx}. Collisions of asymmetric systems is feasible at the BNL RHIC (as demonstrated by the $p$+Au, $d$+Au and $^{3}$He+Au runs~\cite{PHENIX:2018lia}) but not at the CERN LHC in collider mode. However, asymmetric collisions can be performed in fixed-target mode by injecting oxygen-16 and neon-20 ions in the SMOG system of the LHCb detector~\cite{LHCb:2673690} (a small sample of Ne+Pb data was collected in 2013). 

\subsection{Shape evolution along the Samarium isotopic chain}\label{sec:3.2}
Certain isotopic chains in the nuclear chart exhibit strong variations in nuclear shapes. While this occurs mainly away from the stability line, the chain of eight stable samarium isotopes (Sm $Z=62$) features a transition from nearly-spherical to strongly-deformed nuclei with increasing neutron number, e.g. from $^{144}$Sm with $\beta_2\approx0.09$, to $^{154}$Sm with $\beta_2\approx 0.34$, with a change in mass number of only about 7\%. Since the hydrodynamic response is expected to be essentially constant over the isotopic chain, these systems offer a strong lever-arm to probe in detail how the initial condition of QGP responds to varying nuclear shapes, e.g. by predicting the coefficients $b_0$ and $b_1$ in Eq.~\eqref{eq:4} using two Sm isotopes and then make predictions of $\beta_2$ for other isotopes~\cite{Jia:2021tzt}. The $\beta_2$ differences among isotopes can be extracted from ratios of flow observables, as done for the BNL RHIC isobar run. The extracted differences from heavy-ion collisions can be compared with nuclear structure knowledge, to study whether shapes evolve similarly when adding neutrons one-by-one in low-energy experiments and high-energy collisions. We stress that these nuclei have been subject of much investigation at low energy, where their properties are nicely consistent across experiments and theoretical frameworks.

It is worth noting, then, that scanning the Sm isotopic chain in high-energy collisions would provide new experimental insight onto the octupole deformations of such nuclei. As demonstrated by the isobar ratios, nontrivial results are expected. Clear observation of octupole, and potentially hexadecapol deformations for such nuclei would showcase the discovery potential of high-energy nuclear collisions as a tool to observe the manifestations of many-body correlations of nucleons in the ground state of nuclei, in a way that is fully complementary to low-energy structure experiments. In turn, this will provide new experimental constraints to test future \textit{ab initio} calculations of such large and deformed systems.

\subsection{The neutron skin of $^{48}$Ca and $^{208}$Pb in high-energy collisions}\label{sec:3.3}
In low-energy experiments, the neutron skins of $^{48}$Ca and $^{208}$Pb, two doubly-magic nuclei with a considerable neutron excess, have been the subject of much work. Dedicated experiments at Jefferson Lab have been devoted to measuring the neutron skin of these species \cite{PREX:2021umo,CREX:2022kgg}. The measured value for $^{208}$Pb is $\Delta r_{np}=0.28 \pm 0.07$ fm, which is systematically larger than predictions from energy density functional theories. The properties of neutron stars (e.g., the tidal deformability) resulting from such a constraint on the EOS turn out to be slightly at variance with those inferred from pulsar and gravitational wave observations, which has sparked intense debate in the community \cite{Fattoyev:2017jql,Reed:2021nqk}. The neutron skin of $^{48}$Ca is instead more in line with the theoretical expectations. We aim to provide new constraints on the neutron skins of $^{48}$Ca and $^{208}$Pb by utilizing high-energy collisions.

Providing a robust estimate of the neutron skin of $^{48}$Ca in high-energy nuclear collisions is rather straightforward. The isotopic chain of calcium has two doubly-magic nuclei, $^{48}$Ca and $^{40}$Ca. The latter has the same number of protons and neutrons, and its neutron skin is much smaller than that of $^{48}$Ca. However, experiments reveal that $^{48}$Ca and $^{40}$Ca have essentially the same charge radius with a difference less than 0.001~fm~\cite{ca40,GarciaRuiz:2016ohj}, such that neutrons alone determine the differences in size between these two isotopes. As discussed in Sec~\ref{sec:2.4}, heavy-ion collisions allow one to experimentally access differences in the neutron skins between nuclei of similar mass. 
Therefore, if $\Delta r_{np}(^{48}{\rm Ca}) \gg \Delta r_{np}(^{40}{\rm Ca})\approx 0$, collisions of such nuclei could isolate
\begin{equation}
    \Delta r_{np} (^{48}{\rm Ca}) - \Delta r_{np} (^{40}{\rm Ca}) \simeq \Delta r_{np}  (^{48}{\rm Ca}). 
\end{equation}
We estimate that this quantity can be accessed with an uncertainty of about 0.02 fm. Any significant deviations from the expectations of low-energy theories or experiments should be ascribed to the modification of the partonic structure of nucleons in nuclear environment at high energy.

Following this idea, the constraints on neutron skin of $^{208}$Pb could be obtained by comparing data from $^{208}$Pb+$^{208}$Pb with data from $^{197}$Au+$^{197}$Au, as the two species are nearly isobars. Therefore, having such collisions at the same beam energy would allow us to determine the difference $\Delta r_{np. \rm Pb} - \Delta r_{np. \rm Au}$ from observables such as $v_2\{4\}$. This information could be combined with an additional estimate of the neutron skin from a method recently developed by the STAR collaboration \cite{STAR:2022wfe}, also at high energy. This method employs the production of $\rho^0$ mesons in photo-nuclear processes in ultra-peripheral collisions using the newly developed spin interference enabled nuclear tomography. The cross section for $\rho^0$ production in dipole-nucleus scattering contains a coherent component determined by the gluon distribution of the target nucleus. Fits of the coherent diffractive $|t|$ distribution within a Woods-Saxon geometry model in $^{197}$Au+$^{197}$Au collisions lead to $\Delta_{\rm np} (^{197}{\rm Au}) = 0.17\pm 0.03(\mathrm{stat.}) \pm 0.08 (\mathrm{sys})$ fm. This method could be readily applied to other species such as $^{208}$Pb via $^{208}$Pb+$^{208}$Pb collisions. It would measure the neutron skin of $^{208}$Pb with an uncertainty that is similar to or even better than that obtained by the PREX-II experiment. We emphasize that the systematic errors are largely correlated in this technique. The experiment should be able to demonstrate whether the extracted neutron skin difference between $^{208}$Pb and $^{197}$Au is compatible with low energy models and measurements (including PREX-II for Pb~\cite{PREX:2021umo}). We note that a short Pb+Pb collision run at RHIC would be sufficient for this purpose. This is a cost-effective experiment with significant impacts on the nuclear physics community as a whole.

Furthermore, it is worth noting that at the energy reached at the LHC, electro-weak (EW) bosons are abundantly produced in nucleus-nucleus collisions via $u\bar{d}\rightarrow W^+$, $d\bar{u}\rightarrow W^-$, and $q\bar{q}\rightarrow Z$ processes. These real EW bosons probe the weak charge distributions in the heavy-ion initial state, i.e., the sum of weak charge distributions from the two colliding nuclei in the overlap region. Therefore, isobar ratios of $W$ and $Z$ boson yields as a function of centrality may provide direct access to the radial distribution of valence and sea quarks, offering an access route to charge radius, mass radius and thus the neutron skin, and compare with that extracted from the PREX-II analysis of the neutral weak form factor of $^{208}$Pb~\cite{Paukkunen:2015bwa} associated with virtual Z boson. Based on 1~nb$^{-1}$ 5.02 TeV Pb+Pb data (from one month of typical heavy-ion running), we expect each LHC experiment to deliver about 600k W bosons and 20k Z bosons reconstructed in the lepton decay channel~\cite{ATLAS:2019ibd,ATLAS:2019maq}. If isobar or isobar-like collisions, such as $^{40}$Ca+$^{40}$Ca vs. $^{48}$Ca+$^{48}$Ca, or $^{208}$Pb+$^{208}$Pb vs. $^{197}$Au+$^{197}$Au become available at LHC, it will be possible to compute ratios of $W$ and $Z$ boson yields as a function of centrality with a statistical uncertainty of order $1$\% ($\sqrt{1/10000}$). This could be achieved in particular at the high-luminosity LHC, in Run5 or beyond ($>2035$) \cite{Citron:2018lsq,Bruce:2021hii}.

\subsection{Initial conditions of heavy-ion collisions}\label{sec:3.4}
The success of the hydrodynamic framework of heavy-ion collisions enables us today to perform quantitative extractions of the transport properties of the QGP via multi-system Bayesian analyses~\cite{Bernhard:2016tnd,JETSCAPE:2020xug,Nijs:2020ors,Xie:2022ght,JETSCAPE:2022ixz}. A major limitation of such extractions is the lack of precise control on the initial condition of the QGP prior to the hydrodynamic expansion. Insights about the energy deposition from two collided nuclei come from the color glass condensate (CGC) effective theory of high-energy QCD \cite{Gelis:2010nm}. There, for a given boosted nuclear profile described by the \textit{thickness} function  $T=\int \rho(x,y,z) dz$, the average energy density deposited in the transverse plane in the collision of, say, nuclei $A$ and $B$, at the instant immediately after the collision occurs is of the form \cite{Gelis:2019yfm}
\begin{equation}
\label{eq:cgc}
   \langle T^{00} \rangle ~ [{\rm GeV}/{\rm fm}^3] \propto T_AT_B.
\end{equation}
Bayesian analyses of heavy-ion data, while constraining transport properties of the QGP, attempt as well to constrain the initial conditions of the collisions. The prediction of the CGC in Eq.~\eqref{eq:cgc} can be tested via a generic parameterized Ansatz for the energy density, such as the \trento{} Ansatz for the energy density per unit rapidity, $dE/dy ~ [{\rm GeV}/{\rm fm}^2]$, namely $dE/dy \propto (\hat T_A^p + \hat T_B^p)^{1/p}$ \cite{Moreland:2014oya}, or its generalized version, $dE/dy \propto (\hat T_A^p + \hat T_B^p)^{q/p}$ \cite{Nijs:2022rme,Giacalone:2022hnz}, where $\hat T$ represents the thickness function constructed solely from the participant nucleons within the colliding ions.  The values of $p$ and $q$ and other model parameters such as nucleon width $w$ and inter-nucleon minimum distance $d_{\rm min}$can then be inferred from the analysis of high-energy collision data.

However, information about the content of the colliding nuclei, which can not be predicted based on the CGC alone, yields a significant uncertainty in our understanding of the energy deposition itself and, in turn, of the QGP transport parameters resulting from fits to data. One example is provided by $v_3$ in Zr+Zr collisions. If one attempted to reproduce the measured $v_3$ in hydrodynamic calculations without implementing any $\beta_3$ parameter for such a nucleus, one would correct a 10\% enhancement of such an observable in central collisions by biasing the extraction of other QGP transport or initial-state properties dramatically. It is the knowledge of the presence of a large octupole deformation from the isobar ratio $v_{3,\rm Zr}/v_{3,\rm Ru}$, that enable us to avoid biasing the extracted QGP features. Bayesian approaches have not yet systematically explored the impact of nuclear shape and radial distributions. Nuclear structure knowledge should be used systematically as a new lever arm to probe the initial condition of collisions of species that are close in mass, and thus obtain better determinations of the QGP transport coefficients.

\begin{figure}[t]
\includegraphics[width=.8\linewidth]{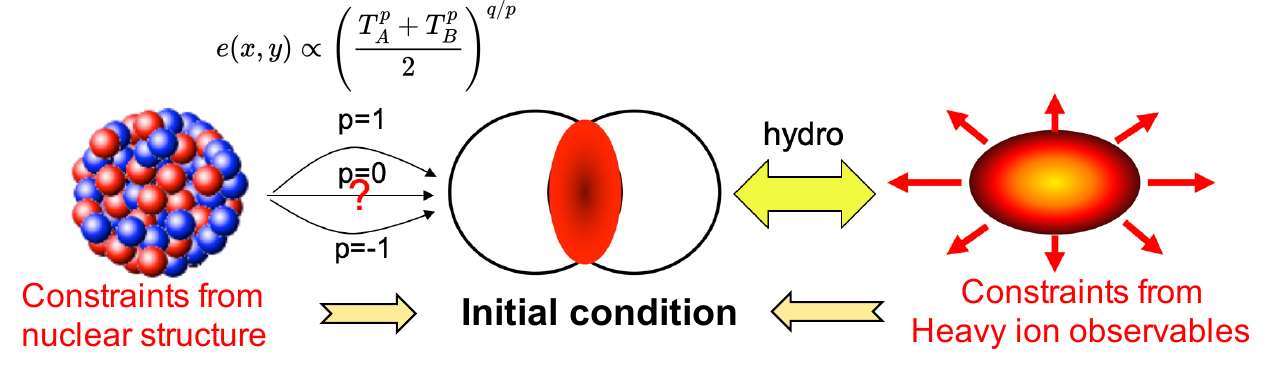}
\caption{\label{fig:6} 
Impact of isobar-like collisions on the goal of the heavy-ion program. Better control on the initial condition can be achieved by exploiting the constraints from both the ratios of final-state observables and the nuclear structure knowledge.}
\end{figure}

As discussed in Sec.~\ref{sec:2.1}, the deviation of isobar ratios from unity probes directly the structural differences between the two species, and the way the initial condition is shaped by two colliding ions. Numerical work shows that the ratios of many observables can be expressed in terms of the differences of Woods-Saxon parameters as a generalization of Eq.~\eqref{eq:5} \cite{Jia:2021oyt},
\begin{equation}
\label{eq:6}
\frac{\mathcal{O}_{{\mathrm Ru}} }{\mathcal{O}_{\mathrm Zr} }\approx 1+ c_0 (R_{0,\rm Ru}-R_{0,\rm Zr})+ c_1 (a_{0,\rm Ru}-a_{0,\rm Zr})+c_2 (\beta_{2,\rm Ru}^2-\beta_{2,\rm Zr}^2)+c_3 (\beta_{3,\rm Ru}^2-\beta_{3,\rm Zr}^2)\;,
\end{equation}
with $\mathcal{O}\equiv p(\nch), v_2, \mbox{or}\; \lr{\pT}$. Crucially, these ratios are insensitive to variations of QGP transport properties~\cite{Zhang:2022fou}. Therefore, the left-hand side of Eq.~\eqref{eq:6} captures the variations of initial conditions in the isobar systems, which are related to the structure parameters on the right-hand side. The coefficients $c_n$ reflect how the initial condition changes when the nuclear structure is varied between the isobars. Via isobar collisions, thus, one can conveniently separate the role played by low-energy nuclear structure input ($R_0$, $a_0$, $\beta_2^2$, $\beta_3^2$) and role played by constraints from the knowledge of high-energy heavy-ion collisions ($c_i$ coefficients)  for observables (isobar ratios) that depend solely on the initial condition of the QGP,  as illustrated in Fig.~\ref{fig:6}.

Collisions from $pp$, $p$+A and A+A have been collected at the BNL RHIC and the CERN LHC. For A+A collisions, we have $^{238}$U+$^{238}$U and $^{197}$Au+$^{197}$Au collisions at the BNL RHIC, and $^{129}$Xe+$^{129}$Xe and $^{208}$Pb+$^{208}$Pb collisions at the CERN LHC. However, none of these pairs are close enough in their mass number~\footnote{Although the 20\% difference in the mass number between $^{238}$U and $^{197}$Au seems not too big, the very large deformation of $^{238}$U makes it non-trivial to precisely constrain the properties of $^{197}$Au (see Ref.~\cite{Giacalone:2021udy} for an attempt).}, which means that the final-state effects do not completely cancel in the ratios~\cite{Giacalone:2017dud}. The model dependencies of these residual effects are then significant enough to preclude a precise extraction of the initial condition. Colliding isobars, or in general species close in mass,  such as $^{197}$Au and $^{208}$Pb, represents an ideal way to constrain the initial condition across the nuclide chart. 

Concerning the possibility of having Pb+Pb reference data at the BNL RHIC,  there are two arguments in the context of hot QCD studies to motivate such an effort, in addition to the neutron skin case pointed out in Sec.~\ref{sec:2.3}. 1) Being doubly magic, $^{208}$Pb is essentially spherical. In contrast, $^{197}$Au has a modest oblate deformation. For the high-precision studies of Au+Au collisions expected from the upcoming sPHENIX program, it would be important, then, to have Pb+Pb collisions as a tool to calibrate the initial condition of Au+Au collisions and ensure that the expectations of the low-energy nuclear theory are compatible with the observations at high energy. 2) Having Pb+Pb systems would also provide a bridge to compare the outcome of Pb+Pb collisions at the BNL RHIC to that of Pb+Pb collisions at the CERN LHC, to study the beam energy dependence of observables. For both these goals, a short Pb+Pb run at the BNL RHIC would be sufficient.

Last but not least, isobar or isobar-like collisions may serve as novel probes of the hard sector, via the analysis of observables such as the production of leading hadron, jets, photons, and heavy flavors. It has already been shown that collective flow of $D$-mesons is sensitive to the deformation of the nucleus \cite{Katz:2019qwv}. Additionally, by constructing ratios of selected observables at a given centrality, or multiplicity, final state effects such as jet quenching are expected to cancel. Deviations from unity in the constructed ratios will provide access to flavor-dependent Nuclear Parton Distribution Function (nPDF), tailored for each underlying hard-scattering process. Interestingly, the precision determination of the impact parameter from bulk particles in coincidence with hard processes means that we can use isobar ratios to detect differences in the transverse spatial distribution of partons at given longitudinal momentum fraction between the two isobars. 

One such example is already discussed in Sec~\ref{sec:3.3} in the context of $W$ and $Z$ bosons for neutron skin measurements. For some of the hard probes, such as high-$\pT$ charged hadrons or inclusive jets, production will be so abundant that even a short run would permit one to determine the isobar ratio with a statistical precision of 1\% or better as a function of centrality, with large cancellation of the systematical uncertainties. One could also study how the isobar ratio evolves with rapidity to detect potential modifications to the nuclear structure inputs due to nPDF or gluon saturation. Isobar ratios of more differential measurements, such as dijet or photon-jet measurements, could probe in more detail the correlation between the final-state medium effects, such as quenching, and the geometry of the hard-scattering processes, such as the path length. For this purpose, collisions of different species should be taken at the same $\sqrt{s_{\rm NN}}$, with similar pileup and detector conditions. Model studies are forthcoming to put this argument on a more quantitative ground.

\subsection{Impact on future experiments: EIC and CBM FAIR}\label{sec:3.5}

Collisions of isobars may provide valuable input to the physics of the planned EIC. One important goal of the EIC program is to understand the partonic structure of nuclei at very high energy \cite{Accardi:2012qut, AbdulKhalek:2021gbh}. At small longitudinal momentum fraction, $x$, the density of gluons may saturate and form the so-called color glass condensate (CGC). EIC will probe gluon saturation using a range of scattering processes in electron-nucleus collisions. In heavy-ion collisions, the modification of parton distributions in nuclei (nPDF) impacts the initial conditions of the QGP, which in turn are imaged via the isobar ratios of bulk and high-$\pT$ observables. In this way, one can gain access to the transverse spatial distribution of partons. Exploiting isobar ratios as a function of rapidity, and in particular as a function of $\sqrtnn$, may provide a unique probe of $x$-dependence of nPDF and gluon saturation. Collisions of the same isobar pair, for example Ru+Ru and Zr+Zr, at different energies could be realized at CBM FAIR~\cite{Almaalol:2022xwv} at $\sqrtnn\leq$ 4.9 GeV, BNL RHIC at $\sqrtnn=200$ GeV, and the CERN LHC at $\sqrtnn>5$ TeV. Any differences between RHIC and LHC in the isobar ratios for soft and hard probes could be used to infer the energy dependence of initial conditions and in turn that of the partonic structure within nuclei. This study will complement the EIC program by providing additional information on the spatial structure of dense gluonic matter. In turn, this information will provide valuable input for the CBM experiment at FAIR in the study of the QCD phase diagram at low temperature and high baryon density~\cite{Almaalol:2022xwv}, in particular, to inform theoretical models such as SMASH~\cite{Weil:2016zrk}, AMPT~\cite{Lin:2021mdn}, and hydrodynamics \cite{Shen:2020mgh, Shen:2017bsr, Spieles:2020zaa, Bluhm:2020mpc, An:2021wof} that aim to describe the dynamical and transport properties of nuclear matter in such conditions.

\section{Brief summary of developments since 2022}\label{sec:3.6}
Since 2022, research devoted to the connection between low-energy nuclear structure and high-energy nuclear collisions has exploded, to the point that it is not possible to cover all the numerous important contributions in this section.

One notable development is the progress made in extracting quantitative information about the structure of nuclei from collider data. Using the Bayesian analysis framework, a recent study demonstrated the possibility of determining the skin of $^{208}$Pb from LHC data \cite{Giacalone:2023cet}, yielding values that align well with low-energy determinations \cite{PREX:2021umo,Hu:2021trw}. A 3+1D simulations of isobar collisions have also revealed the sensitivity to the neutron skin of Ru and Zr nuclei in observables that compare baryon to electric charge stopping~\cite{Pihan:2024lxw}. Additionally, quantitative insights into nuclear deformations were obtained through comparative measurements of $^{238}$U+$^{238}$U and $^{197}$Au+$^{197}$Au collisions conducted by the STAR Collaboration~\cite{STAR:2024eky}. When combined with high-precision hydrodynamic calculations and the relatively well-understood, modest quadrupole deformation of $^{197}$Au \cite{Bally:2023dxi,Ryssens:2023fkv}, these measurements enabled the quantitative extraction of the intrinsic surface deformation parameters, $\beta_{2\rm U}$ and $\gamma_{\rm U}$, of the $^{238}$U nucleus. The results for $\beta_{2\rm U}$ align with low-energy nuclear physics results, while the non-zero value of $\gamma_{\rm U}$ provides evidence for a shape that breaks axial symmetry, shedding light on an aspect of this nucleus that is poorly known in low-energy approaches. 

These studies, among others, suggest a wide range of applications for the high-energy nuclear structure imaging method. Potential applications include: systematically determining structural properties of both even- and odd-mass ground states; probing higher-order deformations such as octupole and hexadecapole shapes~\cite{Ryssens:2023fkv,Xu:2024bdh}; discerning the ``softness'' of the nuclear deformation including the imprint of shape fluctuations or shape coexistence~\cite{Dimri:2023wup}, which could reveal nuclear shape phase transitions~\cite{Zhao:2024lpc};  and utilizing isobar collisions to aid in the search for neutrinoless double beta decay through complementary tests of theory predictions for nuclear matrix elements \cite{Belley:2024zvt,Zhang:2024tzr}. Additionally, progress has been made in formalizing the connection between low-energy theory and high-energy observables based on correlation techniques~\cite{Giacalone:2023hwk}.  The ALICE Collaboration has also performed a measurement of many correlation observables to reveal deformation effects in $^{129}$Xe+$^{129}$Xe collisions~\cite{ALICE:2024nqd}, although this was limited by event statistics.

Extending these investigations to smaller systems, the influence of nuclear structure on collision observables has also been explored. This is motivated by the availability of high-energy $^{16}$O+$^{16}$O collisions at RHIC \cite{Huang:2023viw}, which will also be collected at the LHC in summer 2025. As of October 2024, numerous papers have appeared on nuclear clustering effects in high-energy collisions (e.g. \cite{YuanyuanWang:2024sgp,Zhang:2024vkh,Prasad:2024ahm} for recent works). Comparative studies involving light nuclei, such as $^{16}$O+$^{16}$O versus $^{20}$Ne+$^{20}$Ne collisions~\cite{Ding:2023ibq,Giacalone:2024luz}, and $^{16}$O+Pb versus $^{20}$Ne+Pb collisions~\cite{Giacalone:2024ixe}, predict significant differences beyond model uncertainties. These differences reflect the pronounced structural variations between $^{16}$O and $^{20}$Ne, including potential alpha clustering effects. Interestingly, high-energy electron-isobar collisions offer another avenue to explore spatial distributions and correlations of nucleons in the ground states, albeit involving different types of observables compared to nucleus-nucleus collisions~\cite{Mantysaari:2023qsq,Giacalone:2023hwk,Lin:2024mnj,Magdy:2024thf}.

Furthermore, progress has been made in utilizing nuclear structure to constrain the initial conditions of heavy-ion collisions. Isobar collisions provide a unique opportunity to study the energy dependence and formation mechanisms of these initial conditions~\cite{Li:2022bhl,Bhatta:2023cqf}. Determining the longitudinal structure of QGP has been particularly challenging due to short-range non-flow effects that contaminate direct measurements; these effects arise from sources like resonance decays and jet fragmentation, which are unrelated to the collective flow.  Previous efforts have relied on observables, such as the $r_n$ correlators~\cite{CMS:2015xmx,ATLAS:2017rij}, that do not have a straightforward connection with the longitudinal structure of the initial conditions. Isobar collisions offer a promising solution by allowing us to vary the initial conditions while keeping non-flow effects constant. Due to that, any differences in the longitudinal dependence of observables between isobaric systems can be attributed to changes in their initial conditions. Recent model studies~\cite{Zhang:2024bcb,Jia:2024xvl} demonstrated that this approach enables the complete subtraction of non-flow influences, effectively isolating the longitudinal structure of the harmonic flow across the entire rapidity range.

\section{Summary}
A major goal of the hot QCD program, the extraction of the properties of the quark gluon plasma (QGP), is currently limited by our incomplete understanding of the QGP's \textit{initial condition}, particularly how it forms from colliding nuclei. Our proposal is to use collisions of carefully selected species to precisely assess how variations in nuclear structure affect the initial condition. Combining this approach with detailed measurements of particle correlations in the final state of heavy-ion collisions offers a new method to probe the geometries and spatial correlations of nucleons in atomic nuclei. This will enable us to test utilize predictions from state-of-the-art \textit{ab initio} nuclear structure theories in a novel setup. We encourage the U.S. nuclear physics community to seize this interdisciplinary opportunity by pursuing collisions of strategically chosen species at high-energy colliders.
\begin{itemize}
   \item \textbf{Impact on the hot QCD program.}~Our ability to determine key properties of the QGP from experimental data is limited by our incomplete understanding of its initial conditions immediately after a heavy-ion collision. Colliding nuclear species with significant differences in structural properties provides a new approach to investigate these initial conditions. Specifically, collisions of nuclei that are similar in mass -- such as isobars -- but different  in structure allow us to measure relative changes in observables that are sensitive solely to the QGP initial conditions. These variations stem from "known" structural differences between the species and help us examine precisely how the QGP is formed from the colliding ions. Therefore, future experiments involving isobar collisions with well-known geometries will help reduce uncertainties in determining QGP properties from data.
   \item \textbf{Impact on the nuclear structure program.}~
   Explaining the emergence of nuclei from fundamental theory is a major goal of the nuclear structure program, which can benefit from its synergy with the hot QCD program based on high-energy heavy-ion collisions. Due to the short timescales of the interaction processes, and the \textit{deterministic} nature of the subsequent hydrodynamic evolution from the initial to the final state,  measurements of particle angular correlations in the final states of high-energy collisions are sensitive to many-body correlations of nucleons, such as nuclear deformations, in the colliding nuclei's ground states.  High-energy colliders thus provide a novel tool to unravel strongly correlated nuclear systems and test \textit{ab initio} theories of nuclear structure rooted in QCD.
    \item \textbf{Importance of future collider runs.}~
    
    Collisions of different nuclear species will allow us to utilize and test the predictions of cutting-edge \textit{ab initio} nuclear structure methods while simultaneously reducing the uncertainty in the QGP properties derived from data. It is timely to undertake such interdisciplinary studies in upcoming collider runs. These efforts should focus primarily on the CERN LHC in Run4 and Run5 beyond 2025, but also take advantage of opportunities at the BNL RHIC before it gives way to the electron-ion collider (EIC). A better understanding of the role of nuclear structure in high-energy collisions will enhance hydrodynamic or transport model simulations of collisions at RHIC's BES-II, and at the future CBM experiment at FAIR, where the connection between initial conditions and final states is more involved. Additionally, ensuring the robustness of the low-energy inputs will be valuable for studying the modification of parton distributions within nuclei, as planned at the future EIC.     
\end{itemize}

\section*{Acknowledgments}
We thank the participants of the EMMI Rapid Reaction Task Force \textit{"Nuclear physics confronts relativistic collisions of isobars"} (\url{ https://indico.gsi.de/event/14430/}) for valuable input.  This work is supported by U. S. Department of Energy, Office of Science, Office of Nuclear Physics, under Award or Contract No. DE-SC002418 (JDB), DE-SC0024602 (SH, JJ), DE-SC0004286 (UH), DE-FG02-10ER41666 (CL, WL), DE-SC0013365, DE-SC0024586 and DE-SC0023175 (DL), DE-SC0011088 (YL), DE-AC02-05CH11231 (MP), DE-FG02-89ER40531 (AT), DE-SC0012704 (BS), DE-SC0021969 and DE-SC0024232 (CS),  DE-SC0023861 (JN), DE-FG02-07ER41521 (ZX); by National Science Foundation under grant number OAC-2103680 (JN); by European Union (ERC, Initial Conditions), VILLUM FONDEN with grant no. 00025462, and Danmarks Frie Forskningsfond (YZ); and by FAPESP projects 2017/05685-2, 2018/24720-6, and 2021/08465-9, project INCT-FNA Proc.~No.~464898/2014-5, and CAPES - Finance Code 001 (ML).

\bibliography{main}{}

\end{document}